Classification:

Social Sciences: Mathematical Demography & Biological Sciences: Population Biology

Title:

# Demographic trajectories for supercentenarians


Author: Byung Mook Weon (Corresponding author)

Affiliation: LG.Philips Displays (Company)

Address: 184, Gongdan1-dong, Gumi-city, GyungBuk, 730-702, Republic of Korea

Telephone number: +82-54-460-3424

Fax number: +82-54-461-3901

E-mail address: bmw@lgphilips-displays.com




# Demographic trajectories for supercentenarians


**Abstract**

A fundamental question in aging research concerns the demographic trajectories at the highest ages, especially for supercentenarians (persons aged 110 or more). We wish to demonstrate that the Weon model enables scientists to describe the demographic trajectories for supercentenarians. We evaluate the average survival data from the modern eight countries and the valid and complete data for supercentenarians from the International Database on Longevity (Robine and Vaupel, (2002) *North American Actuarial Journal* 6, 54-63). The results suggest that the Weon model predicts the maximum longevity to exist around ages 120-130, which indicates that there is an intrinsic limit to human longevity, and that the Weon model allows the best possible description of the demographic trajectories for supercentenarians.




## 1. Introduction

Fundamental studies of the aging process have lately attracted the interest of researchers in a variety of disciplines, linking ideas and theories from such diverse fields as biochemistry to mathematics (1). The fundamental model of population biology is the Gompertz model (2), in which the human mortality rate increases roughly exponentially with increasing age at senescence. The Gompertz model is most commonly employed to compare mortality rates between different populations (3). However, no mathematical model so far, including the Gompertz model, has been suggested that can perfectly approximate the development of the mortality rate over the total life span (4). Particularly in modern research findings, it seems to be obvious that the mortality rate does not increase according to the Gompertz model at the highest ages (5-7), and this deviation from the Gompertz model is a great puzzle to demographers, biologists and gerontologists. In order to describe the human demographic trajectories, the Weon model has been proposed by Weon, who has established the concept, model, methodology, generality, definition of maximum longevity and principle in a series of papers (the pre-prints are available at arXiv/q-bio.PE/0402011, 0402013, 0402034, 0403010 and 0403017; the latter two papers are recommended).

The demographic trajectories at the highest ages are a fundamental question for



studying aging and longevity, especially for supercentenarians. For example, some researchers have suggested the mortality curves tending but never reaching a plateau or a ceiling of mortality (8-10), whereas others have suggested that the mortality curves could decrease after having reached a maximum (5-7). Specifically, the mortality curves for higher ages (110+) are essential to understand the maximum longevity for humans. According to the Weon model, the quadratic expression of the age-dependent shape parameter for ages 80-109 is valid with a certain degree of university in many modern developed countries (see arXiv/q-bio.PE/0403010), which enables us to predict that the mortality rate decreases after a plateau around ages 110-115 and the maximum longevity emerges around ages 120-130. If the quadratic expression is still valid for ages 110+, we are able to describe the mortality curves at the highest ages. The trajectories of the mortality dynamics (deceleration, plateau and decrease) at the highest ages by the Weon model seem to be consistent with other assertions (5-7, 11).

In this paper, we wish to evaluate the validity of the Weon model. The aim of this study is to demonstrate whether or not the Weon model enables scientists to describe the demographic trajectories for supercentenarians. The validity can be evaluated by i) the prediction for ages 110+ from the data for ages 80-109, ii) the estimation of the combined data for ages 80-109 and for ages 110+ (supercentenarians)



and iii) the estimation of the only data for ages 110+ (supercentenarians). This study is based upon the number of supercentenarians by Robine and Vaupel (6, 7) and the Weon model most recently proposed by Weon. Firstly we wish to summarize the Weon model and secondly we wish to apply the Weon model to evaluate the average survival data from the modern eight countries and the valid and complete data for supercentenarians from the International Database on Longevity (7). In particular, the age dependence of the shape parameter for ages 80+ inevitably leads to the maximum longevity, which indicates that there is an intrinsic limit to human longevity. All demographic trajectories for human longevity can be explained by the complemetarity between the aging and longevity tendencies.

## 2. Materials and Method

### 2.1. Demographic functions

Three demographic functions are particularly useful (12): the density function, the survival function and the mortality function; Let $f$ be the (probability) density function describing the distribution of life spans in a population. The cumulative density function, $F$, gives the probability that an individual dies before surpassing age $t$ (indicating a continuous variable). The survival function, $s$, gives the complementary



probability ($s = 1 - F$) that an individual is still alive at age $t$. The mortality function, $m$, is defined as the ratio of the density and survival functions ($m = f / s$). Thus, the mortality function gives the probability density at age $t$ conditional on survival to that age. In life table notation, the probability of surviving to age $x$, which indicates a discrete variable, $s$, would be denoted as $l_x / l_0$. The continuous density function, $f$, would be replaced by the discrete function, $d_x$ (or $d_x / l_0$), which gives the number (or population) of life table deaths in age interval from $x$ to $x + 1$. Finally, the mortality function, $m$, would be written in the life table as $m_x$ and is known to demographers and actuaries as the force of mortality.

*2.2. Demographic data*

We use the average survival data for ages 0-109 from the eight countries, Denmark, England & Wales, Finland, France, Japan, Netherlands, Norway and Sweden. Demographic data, the period life tables (for all sexes, 1x1, between 1997-2001), for the eight countries were taken from the Human Mortality Database (13). The eight countries were chosen to compare the IDL (International Database on Longevity) database limited to the nine countries (the eight countries + Belgium) for the valid and complete data for Supercentenarians (110-122 years) (we used the IDL3 lists in ref. 7). We excluded Belgium from the average data, since there was no demographic data for



Belgium in the Human Mortality Database, but we expect that there are little differences

of the average data, regardless of whether the average data include Belgium or not.

## 3. Weon model

### 3.1. Concept

The original concept was obtained as follows: typical human survival curves

show i) a rapid decrease in survival in the first few years of life and ii) a relatively

steady decrease and then an abrupt decrease near death thereafter. Interestingly, the

former behavior resembles the Weibull survival function with $\beta < 1$ and the latter

behavior seems to follow the case of $\beta >> 1$. With this in mind, it could be assumed

that shape parameter is a function of age (see arXiv/q-bio.PE/0402011). It is especially

noted that the shape parameter can indicate a 'rectangularity' of the survival curve. The

reason for this is that as the value of the shape parameter becomes a high value, the

shape of the survival curve approaches a further rectangular shape.

### 3.2. Model

The Weon model is derived from the Weibull model (14) and it is simply

described by two parameters, the age-dependent shape parameter and the characteristic



life. The age-dependent shape parameter enables us to model the survival and mortality functions and it is expressed as follows,

$$s = \exp(-(t/\alpha)^{\beta}) \qquad\qquad [1]$$

$$m = (t/\alpha)^{\beta} \times [\frac{\beta}{t} + \ln(t/\alpha) \times \frac{d\beta}{dt}] \qquad\qquad [2]$$

where $\alpha$ denotes the characteristic life (or the scale parameter, $t = \alpha$ when $s = \exp(-1)$) and $\beta$ denotes the shape parameter as a "function of age". The Weon model is completely different with the Weibull model in the age dependence of the shape parameter. The fact that the shape parameter for humans is a function of age is valid with a certain degree of universality in many countries (see arXiv/q-bio.PE/0402011 and 0403010). The density function by the Weon model can be expressed as the multiplication of the survival and mortality functions by the mathematical relationship of '$f = m \times s$'.

### 3.3. Methodology

We could evaluate the age dependence of the shape parameter to determine an adequate mathematical expression of the shape parameter, after determination of the



characteristic life graphically in the survival curve. Conveniently, the value of the characteristic life is always found at the duration for the survival to be '$\exp(-1)$'; this is known as the characteristic life. This feature gives the advantage of looking for the value of $\alpha$ simply by graphical analysis of the survival curve. In turn, with the observed value of $\alpha$, we can plot the shape parameter as a function of age by the mathematical equivalence of '$\beta = \ln(-\ln s)/\ln(t/\alpha)$'. If $\beta$ is not constant with age, this obviously implies that '$\beta$ is a function of age'.

In empirical practice, we could successfully use a polynomial expression for modeling the shape parameter as a function of age: $\beta = \beta_0 + \beta_1 t + \beta_2 t^2 + \ldots$, where the associated coefficients could be determined by a regression analysis in the plot of shape parameter curve. And thus, the derivative of $\beta$ is obtained as follows: $d\beta/dt = \beta_1 + 2\beta_2 t + \ldots$, which indicates again that the shape parameter for humans is a function of age. Roughly a linear expression is useful for ages 0-80. But for the best fits to the demographic data over the total life span; a cubic or a biquadratic expression is appropriate for ages 0-20, a linear or a quadratic expression is appropriate for ages 20-80 and a quadratic expression is appropriate for ages 80+ (see arXiv/q-bio.PE/0403010).

On the other hand, $\beta$ mathematically approaches infinity as the age $t$ approaches the value of $\alpha$ or the denominator '$\ln(t/\alpha)$' approaches zero. This feature



of $\beta$ can leave 'trace of $\alpha$' in the plot of $\beta$, so we can observe variations of $\beta$ and $\alpha$ at once in the plot of the shape parameters. If $\beta$ (except for the mathematical singularity or trace of $\alpha$) can be expressed by an adequate mathematical function, the survival and mortality functions can be calculated by the mathematically expressed $\beta$. Only two parameters, $\alpha$ and $\beta$, determine the survival and mortality functions.

*3.4. Generality*

The Gompertz model (2) and the Weibull model (14) are the most generally used models at present (15). Interestingly, the Gompertz model is more commonly used to describe biological systems, whereas the Weibull model is more commonly applicable to technical devices (15). In the recent paper (see arXiv/q-bio.PE/0403010), we could see that the traditional models, the Gompertz and Weibull models, may be generalized by the Weon model through the approximate relationship of '$\ln m \propto \beta$' after adulthood (for ages ~20+). The Weon model approximates the Gompertz model when '$\beta \propto t$' and the Weibull model when '$\beta = $ constant'. That is, the Gompertz model is a special case of a linear expression for $\beta$ and the Weibull model is a special case of a constant shape parameter. Particularly, the mortality rate would deviate from the Gompertz model when $\beta$ has a non-linear behavior (before age ~20 or after age ~80).



Thus, $\beta$ is a measure of the deviation from the Gompertz model at higher ages.

*3.5. Definition of maximum longevity*

In general, the term of "*longevity*" means the "*duration of life*". In a sense, the "*maximum longevity*" can be used to mean the "*maximum duration of life*" of a given population. However, what we know is the "*maximum age at death*", which means the oldest age at death observed in a given population during a given time period (16). The Weon model suggests the simple mathematical definitions of the maximum longevity.

In principle, the mortality function should be mathematically positive ($m > 0$). Therefore, the criterion for the mathematical limit of longevity, implying the maximum longevity which is able to be determined by the mortality dynamics in nature, can be given by,

$$\frac{d\beta}{dt} > -\frac{\beta}{t\ln(t/\alpha)} \qquad [3]$$

We successfully used a quadratic expression for the description of the shape parameter after age 80. Interestingly, the quadratic coefficient ($\beta_2$) is important to evaluate the maximum longevity (see arXiv/q-bio.PE/0403010).



On the other hand, perhaps the most common notion of a limit in the study of human longevity is the *limited-life-span hypothesis*, which states that there exists some age $\omega$ beyond which there can be no survivors. This hypothesis can be expressed by any one of the following three formulas (12): " $f = 0$, $s = 0$ or $\lim\limits_{t \to \omega} m = \infty$ $(t \geq \omega)$." By the way, according to the Weon model, the survival function is not zero, although it has extremely low values at the highest ages, but the mortality function can be zero at the maximum longevity (see arXiv/q-bio.PE/0403017). The Weon model suggests that the maximum longevity can be defined as follows: "at $t = \omega$, $f = 0$ and $m = 0$, because of $s \neq 0$." In fact it is possible that the survival rate approaches zero but it is not zero. Even at the maximum longevity ($\omega$), the survival function need not be always zero. Instead, the decrease rate of the survival rate with age ($-ds/dt$; the minus indicates the decrease) should be zero at the maximum longevity. This means that the density function ($f = -ds/dt$) should be zero and thus the mortality function ($m = f/s$) should be zero at the maximum longevity, since the survival function is not zero ($s \neq 0$). Therefore, the maximum longevity can be simply defined as follows,

$$-\frac{ds}{dt} = 0 \qquad\qquad [4]$$



This simple mathematical expression for the maximum longevity makes sense and comprehends the definitions by the density and mortality functions as follows: "at $t = \omega$, $f = 0$ and $m = 0$, because of $s \neq 0$." The values for the maximum longevity calculated from the Eq. [3] and [4] are mathematically identical. In practice, we can identify the maximum longevity at the moment that the survival curve should level in the plot of the survival curves. The mortality curves extrapolated by the Weon model may approach zero at the maximum longevity, which is due to the nature of the density function: That is, the decrease rate of the survival function with age ($-ds/dt$) or the density function should be zero at the maximum longevity (see arXiv/q-bio.PE/0403017).

*3.6. Principle for longevity: complementarity*

The essence of the Weon model is the age dependence of the shape parameter. What is the origin of the age-dependent shape parameter? According to the Weon model, in principle for the highest value of $s$ or for longevity at all times, the shape parameter should be variable according to the characteristic life; "for longevity, $\beta$ increases at $t < \alpha$ but it decreases at $t > \alpha$." This is attributable to the nature of biological systems to strive to survive healthier and longer (see arXiv/q-bio.PE/0403010).



Empirically, the quadratic coefficient ($\beta_2$) indicates the decrease of $\beta$ at $t > \alpha$. Thus, the longevity tends to increase with increasing $\beta_2$. On the other hand, the mortality dynamics (deceleration, plateau and decrease) are a consequence of the decrease of $\beta$ at $t > \alpha$. The quadratic expression is obviously related with the mortality dynamics at $t > \alpha$, which induces the maximum longevity. Interestingly, the maximum longevity tends to decrease with increasing quadratic coefficient ($\beta_2$). It seems that the maximum longevity decreases as the longevity tendency increases, which indicates the "complementarity principle on longevity". It is very interesting that the reason for longevity may be the reason for limit of longevity in nature. Therefore, the age dependence of the shape parameter seems to be governed by the complementarity principle on longevity; "for longevity, $\beta$ increases at $t < \alpha$ but it decreases at $t > \alpha$, resulting in that the maximum longevity decreases as the longevity tendency increases in nature" (see arXiv/q-bio.PE/0403010).

On the other hand, it seems that the density function indicates the effectiveness for longevity between the survival and mortality functions ($f = m \times s$). The mortality rate tends to increase with decreasing the survival rate, which is due to the complementarity between the survival and mortality functions; "For longevity, individuals tend to reduce the mortality rate and strive to improve the survival rate." As



a result, there exists the 'maximum' effectiveness between the survival and mortality rates at the 'characteristic life' (see arXiv/q-bio.PE/0403017).

If that is the case, it can be suggested that there may be two parts of rectangularization for longevity as follows: The survival function is rectangularized by the increase of the shape parameter before characteristic life ( $0 \sim \alpha$ ) – it is the first part. The survival function is rectangularized by the decrease of the shape parameter after characteristic life ( $\alpha \sim \omega$ ) – it is the second part. The first and second parts can be overlapped as one rectangularization as the characteristic life approaches the maximum longevity ( $\alpha \rightarrow \omega$ ) (see arXiv/q-bio.PE/0403017). This paradigm of rectangularization for longevity makes sense and comprehends the conventional paradigm of rectangularization (17, 18).

The principle indicates that there is fundamentally the complemetarity between the aging and longevity tendencies. For longevity, "increasing $\beta$ at $t < \alpha$ but decreasing $\beta$ at $t > \alpha$" or "reducing $m$ and improving $s$" are attributable to the nature of biological systems to strive to survive healthier and longer. The complementarity inevitably leads to the existence of the maximum longevity, which indicates that there is an intrinsic limit to longevity.



## 4. Results

*4.1. Analysis of data up to age 109*

We analyze the characteristic life to be 85.18 years in the average survival curve as shown in Fig. 1 (a). We verify the age dependence of the shape parameter for the average survival data from the eight countries in Fig. 1 (b). Of course, we can see the trace of $\alpha$ in Fig. 1 (b). Especially for ages 80-109, we expect the age dependence of the shape parameter to be a quadratic expression and we obtain a reasonable quadratic fit result in Fig. 1 (c). Through modeling the shape parameter as a quadratic expression for ages 80-109, we inevitably obtain the mortality curve to decelerate, level, decrease and approach zero at a maximum longevity in Fig. 1 (d). In this case, the mortality curve shows that the plateau (or the maximum) is predicted to be "0.62 at 110 years" and the maximum longevity is predicted to be "123.6 years". It is natural that the survival curve should level at the maximum longevity in Fig. 2 (a), through the definition of the maximum longevity "$-ds/dt = 0$".

*4.2. Analysis of data including supercentenarians*

We predict the average survival rate to be "$5.278 \times 10^{-5}$" at age 110 from the data for 80-109 in Fig. 2 (a). Assuming the continuity of the survival curve, we are able



to adjust the number of supercentenarians from the IDL to be "$5.278 \times 10^{-5}$" at age 110. This adjustment result is seen in Fig. 2 (a) (triangle). From the combined data for ages 80-109 from the eight countries and for ages 110+ from the IDL, we are able to evaluate the age dependence of the shape parameter in Fig. 2 (b). In this case, we verify that the quadratic expression is absolutely valid for the shape parameter at the highest ages. The consequential trajectories of the survival and mortality curves will show that they should have a maximum longevity. For example, the mortality curve is predicted to decelerate, level, decrease and approach zero at a maximum longevity in Fig. 2 (c). In this case, the mortality curve shows that the plateau (or the maximum) is predicted to be "0.62 at 110 years" and the maximum longevity is predicted to be "123.0 years", which is very similar to the predicted from ages 80-109. Therefore, it seems that the survival and mortality curves evaluated for ages 80-109 is appropriate for the prediction for ages 110+.

## 5. Discussion

This above analysis is based upon an assumption of the continuity of the survival curve; that is, the survival curves will continue over the total life span to follow a reasonable mathematical model. At this moment, we wish to discuss the validity of



our assumption.

What does the adjusted value at age 110 mean? In the above analysis, the adjusted value for age 110 is estimated to be "$5.278 \times 10^{-5}$", which is extrapolated by the survival curve through modeling the shape parameter for ages 80-109. It should be noted that this value indicates the possibility of persons alive at age 110 in the population of the modern eight countries. This adjusted value at age 110 may indicate that there are perhaps "~5.3" persons in the population of 100,000 or "~53 per million" at age 110. Is this reasonable? However, the valid and complete number of supercentenarians in the literature (7) seems to be fewer than the expected number. The valid and complete number of persons at age 110 from the nine countries (Belgium, Denmark, England & Wales, Finland, France, Japan, Netherlands, Norway and Sweden) is "82" according to the IDL (7), which is "~0.28 per million", considering that the total population in the nine countries is ~300 millions (19). There is a significant gap (~200 times) between the expected population (~53 per million) and the observed population (~0.28 per million) at age 110 as shown in Fig. 2 (d).

At this moment, the adjusted value at age 110 gives us an important insight into the number of supercentenarians and the maximum longevity. i) For the case of ~53 per million (adjustment), we have to re-estimate the number of supercentenarians. That is,



there may be more supercentenarians in the nine countries. Supercentenarians first emerged consistently in the 1960s and their numbers have been expanding dramatically since (7). Nevertheless, we expect that the number or the probability at age 110 will not surpass the value to be expected at age 110 (~53 per million), assuming the continuity of the survival curve. In spite of the dramatic increase of the number for supercentenarians, the Weon model suggests that there will exist the maximum longevity around ages 120-130 as shown in Fig. 2 (c). ii) For the case of ~0.28 per million (no adjustment), we have to solve the following question – The average survival probability at age 109 from the eight countries is ~80 per million. Then, the only "~0.4 percent" of survivors at age 109 can be alive at age 110, which indicates that most of survivors alive at age 109 should die at age 110. Is this realistic? Why most of survivors at age 109 should die at age 110? At present we have no answer for this question about the population of supercentenarians. Nevertheless, we can assess the applicability of the Weon model to describe the demographic trajectories for supercentenarians. For the case of ~0.28 per million (no adjustment), we evaluate again the shape parameter pattern for ages 110-122 with no adjusted (raw) data in Fig. 2 (e) and finally we obtain the survival curve for ages 110-122 in Fig. 2 (f) (below). In this case, we can see that the survival curve decreases with age and levels at the maximum longevity around 120



years. As a result, the Weon model allows the best possible description of the demographic trajectories for supercentenarians through modeling the age-dependent shape parameter.

Finally on all occasions, the results suggest that the Weon model predicts the maximum longevity to exist around ages 120-130, which indicates that there is an intrinsic limit to human longevity, and that the Weon model allows the best possible description of the demographic trajectories for supercentenarians.

**Figure Legends**

**Fig. 1.** Demographic trajectories for ages 0-109. (a) The average survival curve for the eight countries and the characteristic life observed to be 85.18 years. (b) The shape parameter as a function of age for the eight countries. (c) Modeling the shape parameter as a quadratic expression between ages 80-109 for the eight countries. (d) Modeling the mortality curve through modeling the shape parameter for ages 80-109 for the eight countries.

**Fig. 2.** Demographic trajectories for ages 110+. (a) Modeling the survival curve through modeling the shape parameter for ages 80-109 for the eight countries and adjustment of IDL data at age 110. (b) Modeling the shape parameter for combined data of the eight countries and the IDL data by adjustment. (c) Modeling the mortality curve through modeling the shape parameter for combined data of the eight countries and the IDL data by adjustment. (d) Modeling the survival curve through modeling the shape parameter for ages 80-109 for the eight countries and the IDL data calculated without adjustment. (e) Modeling of the shape parameter for ages 110-122 for the IDL data calculated without adjustment. (f) Modeling the survival curve through modeling of the shape parameter for ages 110-122 for the IDL data calculated without adjustment.



**Acknowledgements**

The author thanks to the Human Mortality Database (Dr. John R. Wilmoth, as a director, in The University of California, Berkeley and Dr. Vladimir Shkolnikov, as a co-director, in The Max Planck Institute for Demographic Research) for allowing anyone to access the demographic data for research.



**Figures**

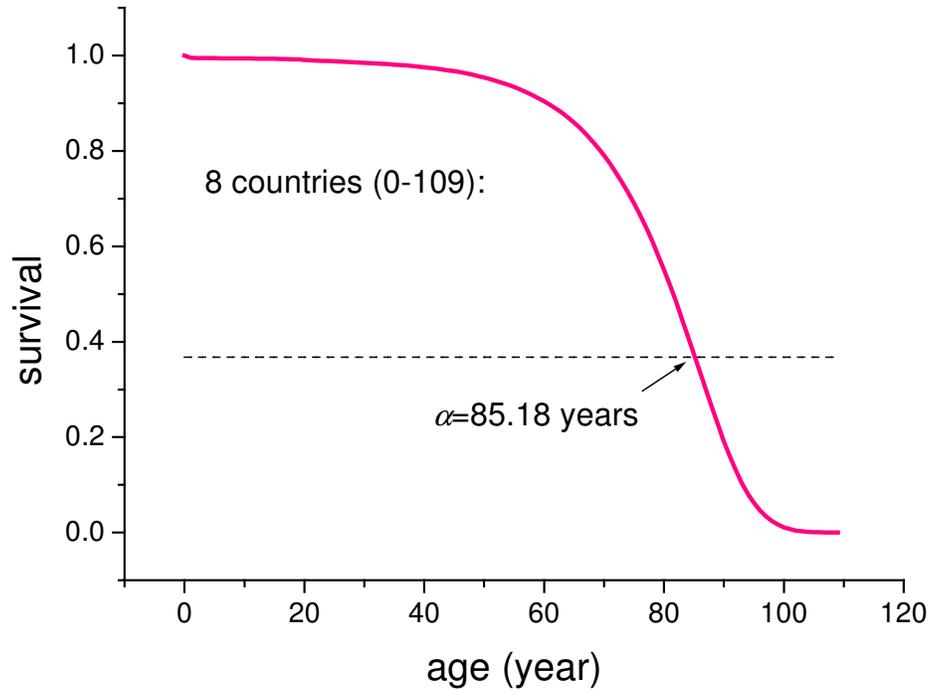

Fig. 1. (a)



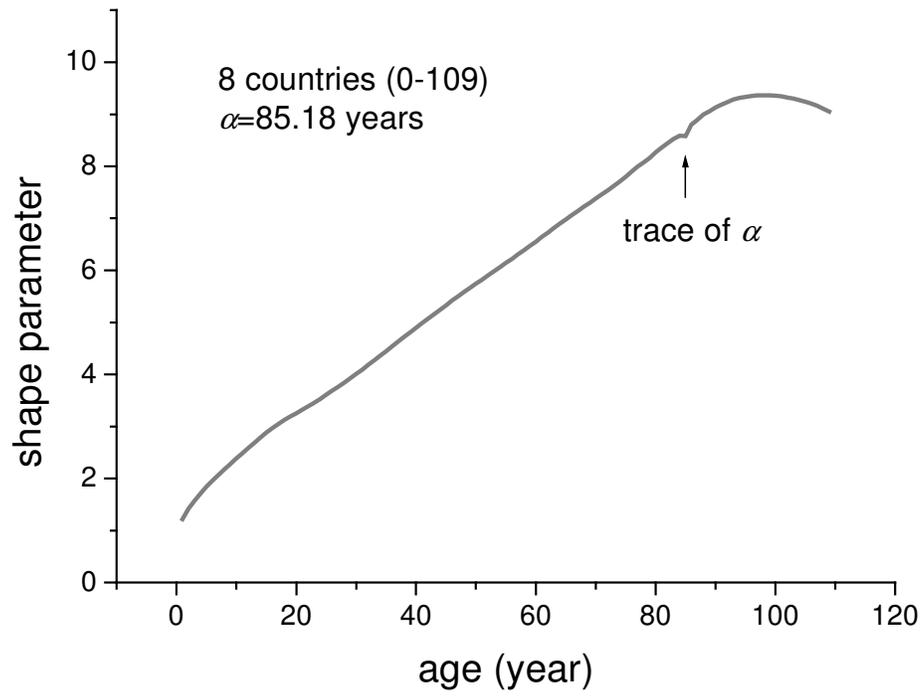

Fig. 1. (b)



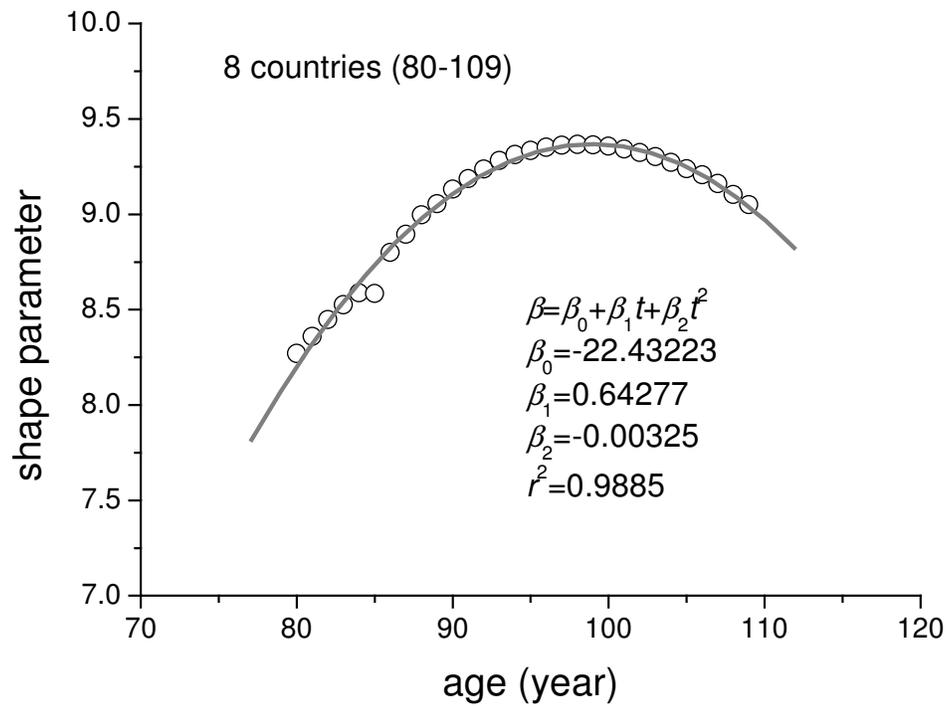

Fig. 1. (c)



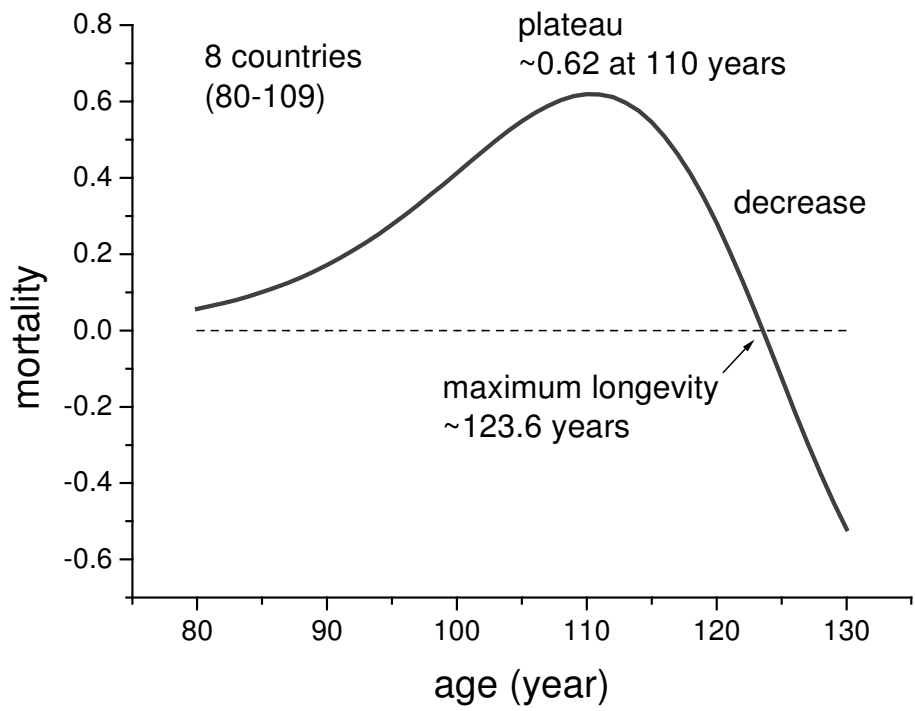

Fig. 1. (d)



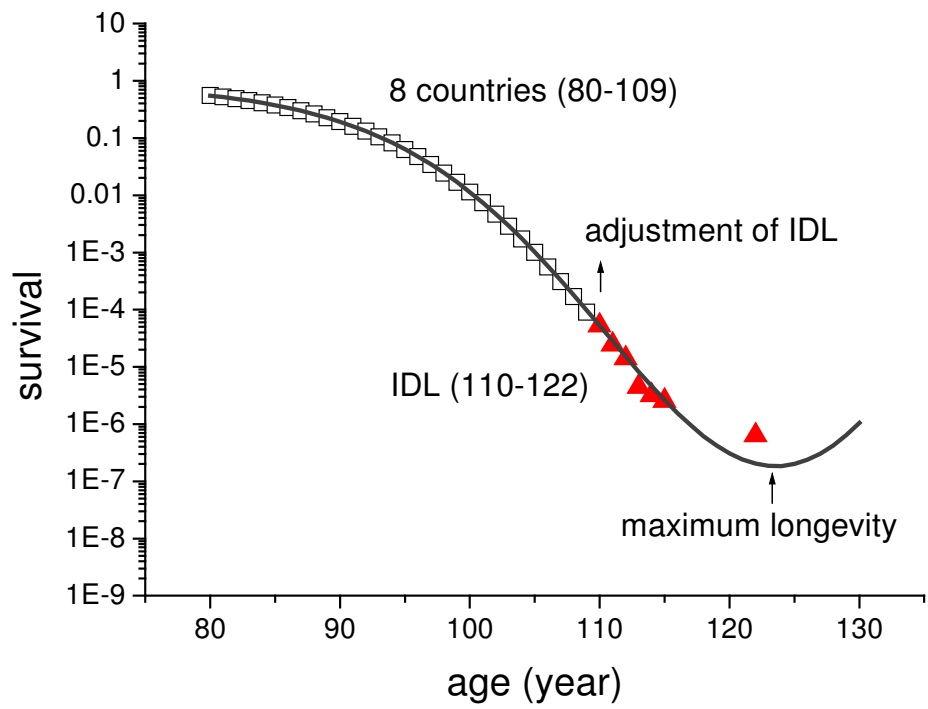

Fig. 2. (a)



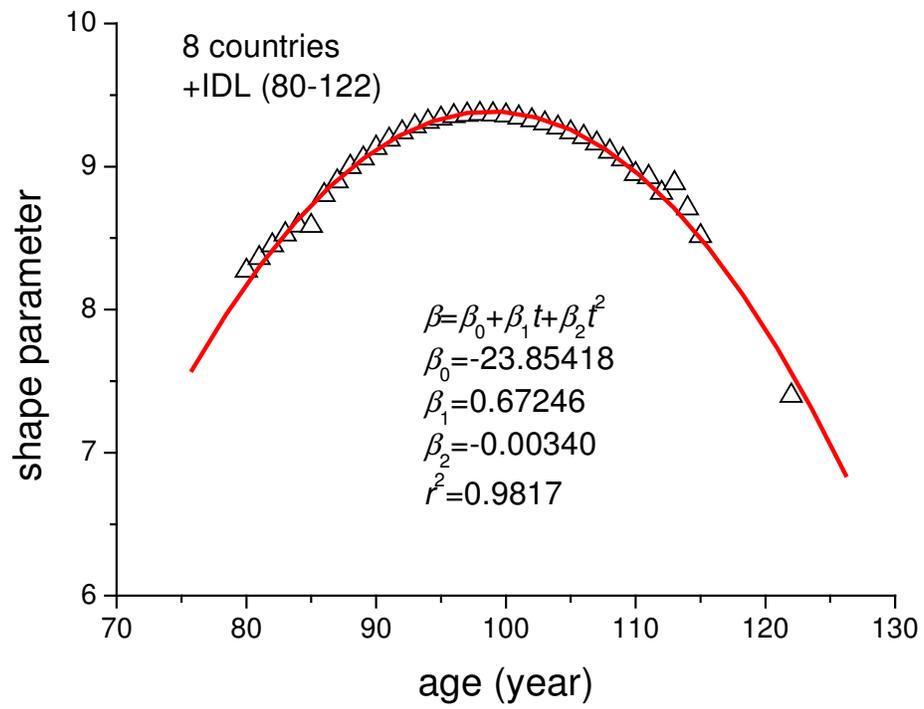

Fig. 2. (b)



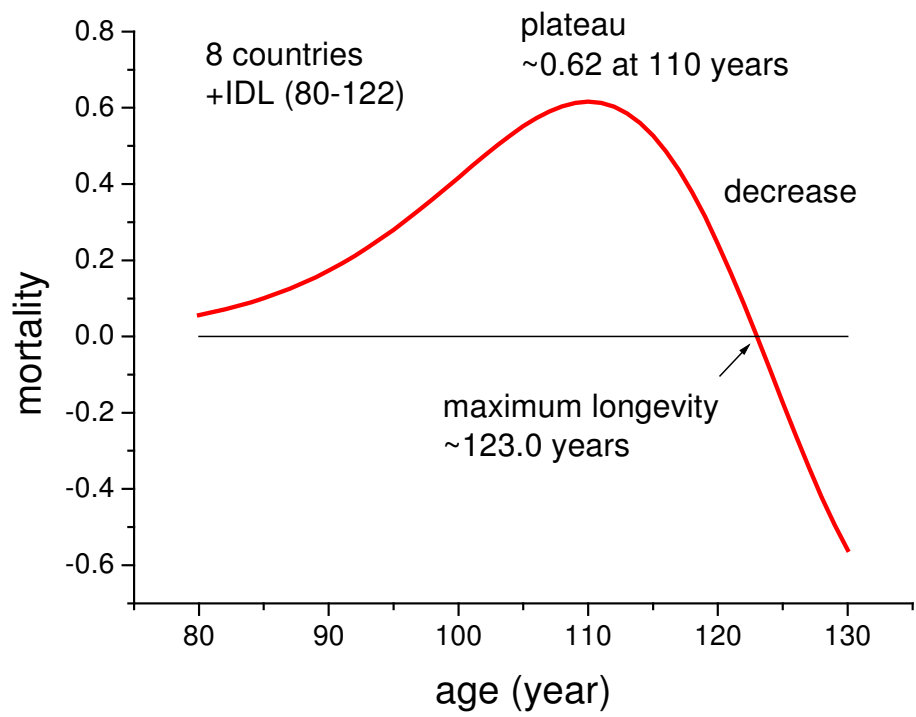

Fig. 2. (c)



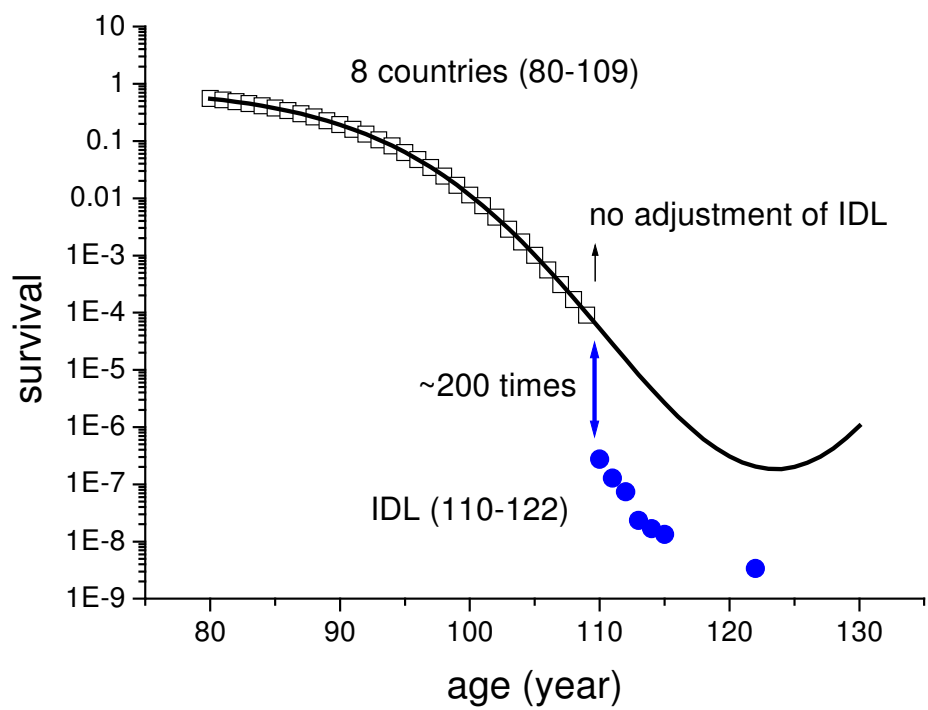

Fig. 2. (d)



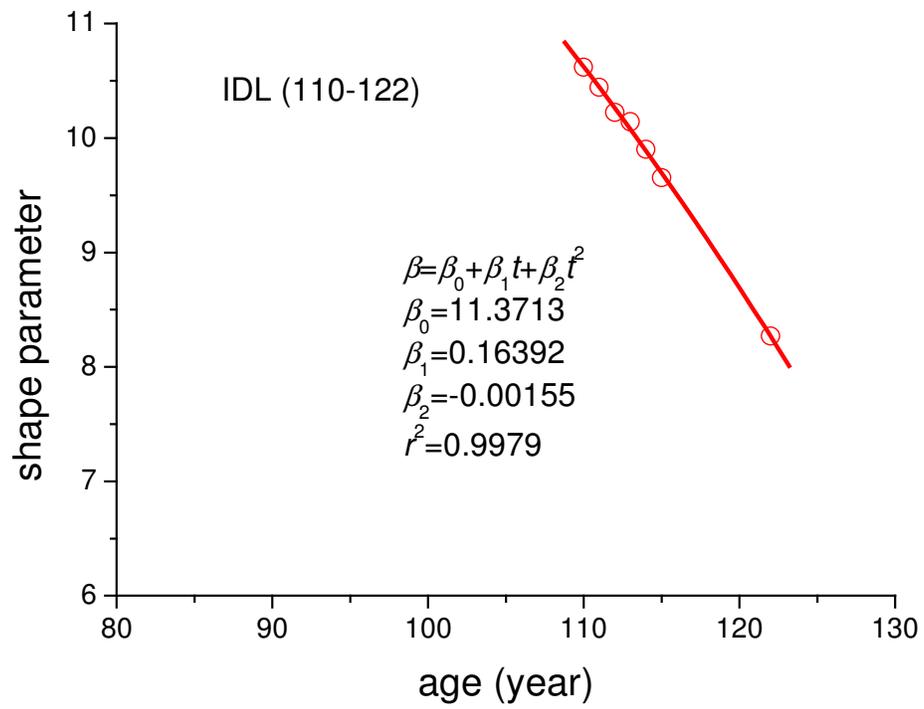

Fig. 2. (e)



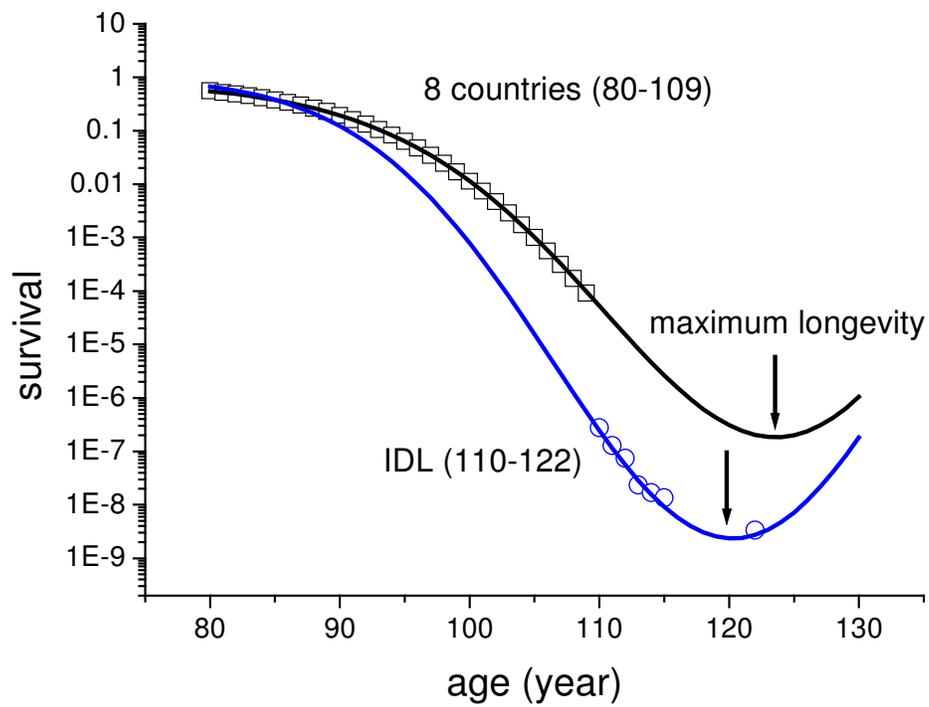

Fig. 2. (f)